\documentclass[technote,9pt]{IEEEtran}
\IEEEoverridecommandlockouts
\usepackage{amssymb}
\usepackage{algorithm}
\usepackage{algorithmic}
\usepackage{amsfonts}
\usepackage{amsmath}
\usepackage{amssymb}
\usepackage{amsthm}
\usepackage{array}
\usepackage{bbding}
\usepackage{bigints}
\usepackage{booktabs}
\usepackage{cite}
\usepackage{cleveref}
\usepackage{color}
\usepackage{diagbox}
\usepackage{epsfig,latexsym}
\usepackage{epstopdf}
\usepackage{graphicx}
\usepackage{fancyhdr}
\usepackage{float}
\usepackage{flushend}
\usepackage{indentfirst}
\usepackage{lastpage}
\usepackage{makecell}
\usepackage{mathtools}
\usepackage{multirow}
\usepackage{pifont}
\usepackage{psfrag}
\usepackage{setspace}
\usepackage{stfloats}
\usepackage{subfloat}
\usepackage{tikz}
\usepackage{times}

\theoremstyle{remark}

\newtheorem{corollary}{\quad \textbf{Corollary}}

\allowdisplaybreaks[4]

\begin{document}
\title{Performance Analysis of Active RIS-aided Systems in the Face of Imperfect CSI and Phase Shift Noise}

\author{Qingchao Li, Mohammed El-Hajjar, \textit{Senior Member, IEEE}, Ibrahim Hemadeh, \textit{Member, IEEE}, Deepa Jagyasi, Arman Shojaeifard, \textit{Senior Member, IEEE}, Lajos Hanzo, \textit{Life Fellow}

\thanks{This work was supported by InterDigital. \textit{(Corresponding author: Lajos Hanzo.)}

Qingchao Li, Mohammed El-Hajjar and Lajos Hanzo are with the Electronics and Computer Science, University of Southampton, Southampton SO17 1BJ, U.K. (e-mail: Qingchao.Li@soton.ac.uk; meh@ecs.soton.ac.uk; lh@ecs.soton.ac.uk).

Ibrahim Hemadeh, Deepa Jagyasi, Arman Shojaeifard are with InterDigital, London EC2A 3QR, U.K. (e-mail: Ibrahim.Hemadeh@InterDigital.com; Deepagurmukhdas.Jagyasi@InterDigital.com; Arman.Shojaeifard@InterDigital.com).}}

\maketitle

\begin{abstract}
The linear minimal mean square error (LMMSE) estimator for active reconfigurable intelligent surface (RIS)-aided wireless systems is formulated. Furthermore, based on the moment-matching method, we employ the Gamma distribution to approximate the distribution of the instantaneous received signal-to-interference-plus-noise ratio (SINR), and then derive the closed-form outage probability and ergodic channel capacity in the presence of realistic channel estimation errors, the thermal noise of RIS amplifiers and the RIS phase shift noise. Our theoretical analysis and simulation results show that the introduction of RIS amplifiers is equivalent to increasing of the transmit power, and also present the performance degradation resulting from the channel estimation error and the RIS phase noise.
\end{abstract}

\section{Introduction}
Reconfigurable intelligent surfaces (RIS) are composed of reflecting elements, each of which can be electronically tuned to adjust the phase shift of impinging signals \cite{basar2019wireless}. With the aid of RIS deployment between the transmit and receiver nodes, the transmission reliability can be enhanced by appropriately configuring the phase shift of each RIS element \cite{li2022reconfigurable,li2022reconfigurable_iot,pang2021irs,pang2021uav}. Therefore, the RIS can also be used for localization \cite{ozturk2022impact}.

Given these benefits, sophisticated mathematical tools have been used for characterizing the theoretical performance analysis of RIS-aided wireless communications \cite{basar2019wireless,yang2020coverage,van2020coverage,yang2020accurate,yang2021performance}. In \cite{basar2019wireless}, Basar \textit{et al.} derived the theoretical channel power gain of single-input-single-output (SISO) RIS-aided systems, which reveals that the received power is proportional to the square of the number of RIS elements, when the direct links is blocked. Based on the central-limit-theorem (CLT), the authors demonstrated that the instantaneous signal-noise-ratio (SNR) approximately follows the non-central chi-square distribution, if the transmitter-RIS link and the RIS-receiver link exhibit Rayleigh fading and the number of reflecting elements is high enough. Based on the CLT harnessed for approximating the distribution of instantaneous SNR in \cite{yang2020coverage}, the outage probability was derived as a function of the cumulative distribution function (CDF) of the instantaneous SNR, while the upper and lower bound of the average channel capacity were derived based on the mean and variance of the instantaneous SNR. In \cite{van2020coverage}, Chien \textit{et al.} employed the Gamma distribution for approximating the instantaneous received SNR, based on which both the coverage probability and the ergodic capacity were theoretically derived. In \cite{yang2020accurate}, Yang \textit{et al.} employed the general-$K$ distribution to approximate the received SNR, which exhibits higher accuracy than the CLT based method. In \cite{yang2021performance}, the closed-form expressions of the outage probability, bit error ratio, and average capacity were derived for the RIS-aided wireless communications, under the assumption of realistic channel state information (CSI) acquisition.

However, the theoretical analysis in the above treatises have the following limitations. Firstly, they were limited to passive RIS-aided wireless communications, where the RIS reflecting elements can only configure the phase of impinging signals without signal amplification. The signals suffer from the twin-hop path-loss of the transmitter-RIS and the RIS-receiver links, which results in low received signal power. To circumvent these limitations, the concept of active RIS is investigated in \cite{jin2021channel,long2021active,zhi2022active}. In the literature two active RIS models have been considered: one equipped with active channel sensors having signal precessing capabilities \cite{jin2021channel}, and another equipped with power amplifiers having no signal processing capabilities \cite{long2021active,zhi2022active}. In this paper we consider using active RIS with no signal processing capabilities, employing power amplifiers that can allow different amplitude gains for different RIS elements. Secondly, the theoretical analysis in above papers is based on the assumption of perfect RIS phase shift design without considering the RIS phase shift noise, which is unrealistic in practical RIS-aided systems having realistic phase shift noise \cite{badiu2019communication,qian2020beamforming,papazafeiropoulos2021intelligent}. To deal with the above issues, our contributions in this compact paper are as follows:

\begin{itemize}
  \item We develop the LMMSE channel estimator for active RIS-aided wireless communications, where the channel's covariance matrix is derived by considering the thermal noise of RIS amplifiers and the RIS phase noise following both the von Mises distribution and the uniform distribution \cite{badiu2019communication,qian2020beamforming,papazafeiropoulos2021intelligent}. This is the first paper considering the channel estimation of active RISs having no signal processing capabilities.
  \item We present the theoretical analysis of the active RIS-aided wireless systems. Specifically, the moment matching method is invoked for approximating the distribution of the instantaneous received signal to interference plus noise ratio (SINR). Then, we present the closed-form outage probability and ergodic channel capacity, taking into account the effect of the thermal noise of RIS amplifiers, the RIS phase shift noise and the channel estimation errors.
\end{itemize}

\textit{Notations:} $\jmath=\sqrt{-1}$. Vectors and matrices are denoted by boldface lower and upper case letters, respectively. $(\cdot)^{\text{T}}$, $(\cdot)^{*}$, and $(\cdot)^{\text{H}}$ represent the operation of transpose, conjugate and hermitian transpose, respectively. $|a|$ and $\angle{a}$ represent the amplitude and angle of the complex scalar $a$, respectively. $\mathbb{C}^{m\times n}$ denotes the space of $m\times n$ complex-valued matrices. $a_n$ represents the $n$th element in vector $\mathbf{a}$. $\mathbf{0}_{N}$ is the $N\times1$ zero vector. $\mathbf{I}_{N}$ and $\mathbf{O}_{N}$ represents the $N\times N$ identity matrix and zero matrix, respectively. $\text{diag}\left\{\mathbf{a}\right\}$ denotes a diagonal matrix with the diagonal elements being the elements of $\mathbf{a}$ in order. $\mathcal{CN}(\boldsymbol{\mu},\mathbf{\Sigma})$ is a circularly symmetric complex Gaussian random vector with the mean $\boldsymbol{\mu}$ and the covariance matrix $\mathbf{\Sigma}$. $\mathbb{E}[\mathbf{x}]$ and $\mathbb{V}[\mathbf{x}]$ represent the mean and the variance of the random vector $\mathbf{x}$, respectively. The covariance matrix between the random vectors $\mathbf{x}$ and $\mathbf{y}$ is denoted as $\mathbf{C}_{\mathbf{x}\mathbf{y}}$. Finally, $\mathop{\sum_{(n_1,n_2,\cdots,n_q)=1}^{N}}$ represents $\mathop{\sum_{n_1=1}^{N}\sum_{n_2=1}^{N}\cdots\sum_{n_q=1}^{N}}\limits_{n_1\neq n_2\neq\cdots\neq n_q}$.

\section{System Model}\label{System_Model}
The RIS-aided wireless communication system model of \cite{zhi2022active} is shown in Fig. \ref{Fig_system_model_active_RIS}, including a single-antenna transmitter, a single-antenna receiver and a RIS having $N=N_x\times N_y$ elements{\footnote{In this work, we focus on the SISO system model in order to provide the closed-form theoretical performance analysis. The investigation of the multi-cell multiple input multiple output (MIMO) system is considered as part of our future work.}, where $N_x$ and $N_y$ represent the numbers of reflecting elements in the horizontal and vertical direction, respectively.

\subsection{Channel Model}
We assume that the direct transmitter-receiver link is blocked and only the RIS-aided two-hop link supports signal propagation. In the transmitter-RIS link (RIS-receiver link), we denote the large scale and small scale fading by $\varrho_{\text{t}}$ ($\varrho_{\text{r}}$) and $\mathbf{g}_\text{t}\in\mathbb{C}^{N\times1}$ ($\mathbf{g}_\text{r}^{\text{H}}\in\mathbb{C}^{1\times N}$), respectively. The large scale fading is given by $\varrho_{\text{t}}=\text{C}_0d_{\text{t}}^{-\alpha_{\text{t}}}$ and $\varrho_{\text{r}}=\text{C}_0d_{\text{r}}^{-\alpha_{\text{t}}}$, where $\text{C}_0$ is the path loss at the reference distance of 1 meter, $d_{\text{t}}$ and $\alpha_{\text{t}}$ denotes the distance between the transmitter and the RIS as well as the corresponding path loss exponent. Furthermore, $d_{\text{r}}$ and $\alpha_{\text{r}}$ represent the distance between the RIS and the receiver as well as the corresponding path loss exponent \cite{long2021active}. In terms of the small scale fading, we assume that both $\mathbf{g}_{\text{t}}$ and $\mathbf{g}_{\text{r}}$ obey Rician fading, given by \cite{han2019large}
\begin{align}\label{Channel_Model_2}
    \mathbf{g}_{\text{t}}\sim\mathcal{CN}(\sqrt{\frac{\kappa_{\text{t}}\varrho_{\text{t}}}
    {1+\kappa_{\text{t}}}}
    \overline{\mathbf{g}}_\text{t},\frac{\varrho_{\text{t}}}{1+\kappa_{\text{t}}}\mathbf{I}_N),
\end{align}
\begin{align}\label{Channel_Model_3}
    \mathbf{g}_{\text{r}}^{\text{H}}\sim\mathcal{CN}(\sqrt
    {\frac{\kappa_{\text{r}}\varrho_{\text{r}}}
    {1+\kappa_{\text{r}}}}\overline{\mathbf{g}}_\text{r}^{\text{H}},\frac{\varrho_{\text{r}}}
    {1+\kappa_{\text{r}}}\mathbf{I}_N),
\end{align}
where $\kappa_{\text{t}}$ and $\kappa_{\text{r}}$ denote the Rician factors, $\overline{\mathbf{g}}_\text{t}$ and $\overline{\mathbf{g}}_\text{r}$ represent the LoS component vectors, each element of which has a unit-modulus and its phase depends on the angle-of-arrival, the angle-of-departure at the RIS and the wavelength $\lambda$.

Let us denote the cascaded channel of $\mathbf{g}_{\text{t}}$ and $\mathbf{g}_{\text{r}}^{\text{H}}$ by $\mathbf{h}$, which is given as follows
\begin{align}\label{Channel_Model_4}
    \mathbf{h}=[g_{\text{t},1}g_{\text{r},1}^{*},g_{\text{t},2}g_{\text{r},2}^{*},\cdots,
    g_{\text{t},N}g_{\text{r},N}^{*}]^{\text{T}},
\end{align}
Given the absence of signal processing capabilities at the RIS elements, the channels $\mathbf{g}_{\text{t}}=[g_{\text{t},1},g_{\text{t},2},\cdots,g_{\text{t},N}]^{\text{T}}$ and $\mathbf{g}_{\text{r}}=[g_{\text{r},1},g_{\text{r},2},\cdots,g_{\text{r},N}]^{\text{T}}$ cannot be estimated separately, and only their cascaded channel $\mathbf{h}$ in (\ref{Channel_Model_4}) can be acquired. Fortunately, estimating the cascaded channel is sufficient for designing the RIS phase shift matrix for data transmission without loss of optimality \cite{liu2021cascaded}.

\begin{figure}[!t]
    \centering
    \includegraphics[width=2.2in]{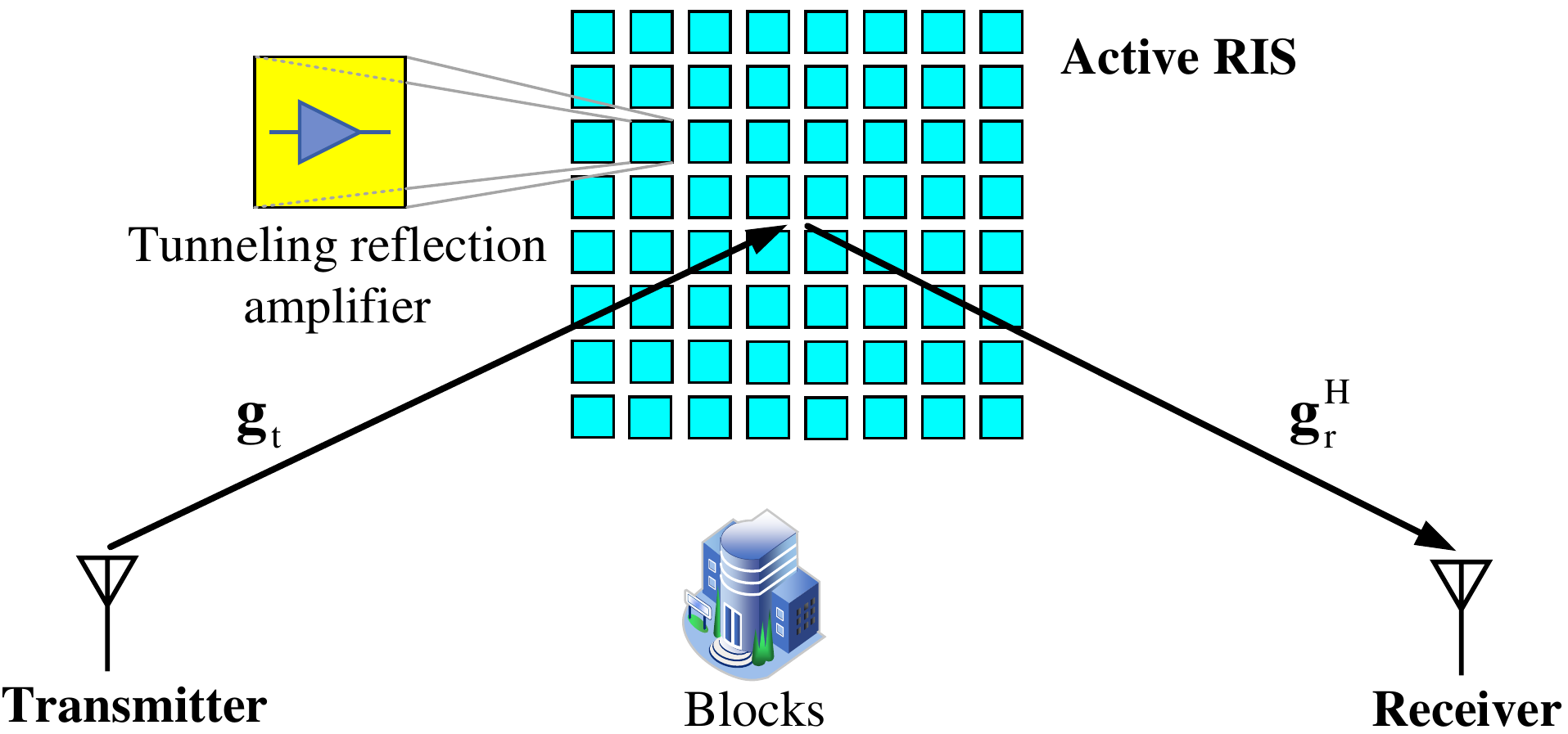}
    \caption{System model of the considered active RIS-aided wireless communication system.}\label{Fig_system_model_active_RIS}
\end{figure}

\subsection{RIS Architecture}\label{System_Model_RIS_Architecture}
The response of the active RIS elements is given by \cite{long2021active}
\begin{align}\label{RIS_Architecture_1}
    \mathbf{z}=[\beta_1\text{e}^{\jmath\theta_1},\beta_2\text{e}^{\jmath\theta_2},\cdots,
    \beta_N\text{e}^{\jmath\theta_N}],
\end{align}
where $\beta_n$ and $\theta_n$ represents the amplitude gain and phase shift of the signals impinging on the $n$th RIS reflecting element. In the passive RIS $\beta_n$ is usually fixed to 1, while in the active RIS $\beta_n$ can be higher than 1 by harnessing a tunneling reflection amplifier for each RIS element \cite{long2021active}. However, since the power amplifiers are active, the thermal noise from RIS reflection amplifiers is inevitable. The power assigned to the transmitter and to the RIS elements, i.e. the values of $\rho,\beta_1,\beta_2,\cdots,\beta_N$, should be optimized under a fixed power budget. In this paper, we assume that $\rho,\beta_1,\beta_2,\cdots,\beta_N$ are prior given, and the power allocation schemes can be found in \cite{long2021active,zhi2022active}. In most treatises, it was assumed that the phase shift can be perfectly configured \cite{basar2019wireless,li2022reconfigurable,yang2020coverage,van2020coverage,yang2020accurate,
yang2021performance,long2021active,zhi2022active}. Due to the realistic RIS hardware impairments, the phase shift of each reflecting element is practically modelled as $\theta_n=\overline{\theta}_n+\tilde{\theta}_n$, where $\overline{\theta}_n$ represents the expected phase shifts, and $\tilde{\theta}_n$ is the phase noise of the $n$th element \cite{badiu2019communication,qian2020beamforming,papazafeiropoulos2021intelligent}. The phase noise $\tilde{\theta}_n$ obeys identically and independently distributed (i.i.d.) random variables having the mean of 0, and it may also be modelled by the von Mises distribution or the uniform distribution \cite{badiu2019communication,qian2020beamforming,papazafeiropoulos2021intelligent}. These may be represented as $\tilde{\theta}_n\sim \mathcal{VM}(0,\varsigma_\text{p})$ and $\tilde{\theta}_n\sim \mathcal{UF}(-\iota_\text{p},\iota_\text{p})$, respectively, where $\varsigma_\text{p}$ is the concentration parameter of the von Mises distributed variables and $(-\iota_\text{p},\iota_\text{p})$ is the support interval of the uniformly distributed variables.

\section{LMMSE Channel Estimation}\label{LMMSE_Channel_Estimation}
In each coherence time, $T$ symbol intervals are employed for estimating the instantaneous channel state information of the cascaded channel $\mathbf{h}$ in (\ref{Channel_Model_4}). Since there is only a single transmit antenna, the pilot symbols transmitted in these $T$ symbol intervals can be identical. For simplicity, we opt for 1. Furthermore, $T$ RIS training patterns, denoted by $\mathbf{z}_1,\cdots,\mathbf{z}_T$, are activated in these $T$ symbol intervals, where $\mathbf{z}_t=[\beta_1\text{e}^{\jmath\theta_{t,1}},
\cdots,\beta_N\text{e}^{\jmath\theta_{t,N}}]$ with $\theta_{t,n}=\overline{\theta}_{t,n}+\widetilde{\theta}_{t,n}$. The signals received in these $T$ symbol intervals are denoted as $y_1,\cdots,y_T$, the thermal noise of the active reflection amplifier at the $t$th symbol interval is denoted as $\mathbf{a}_t=[a_{t,1},\cdots,a_{t,N}]^\mathrm{T}$ with $\mathbf{a}_t\sim\mathcal{CN}(\mathbf{0}_N,\sigma_a^2\mathbf{I}_N)$, and the additive noise at the receiver is $\mathbf{w}=[w_1,\cdots,w_T]^\mathrm{T}$ with $\mathbf{w}\sim\mathcal{CN}(\mathbf{0}_T,\sigma_w^2\mathbf{I}_T)$. Then, we can arrive at
\begin{align}\label{Channel_Estimation_1}
    \mathbf{y}=\sqrt{\rho\varrho_\text{t}\varrho_\text{r}}\mathbf{Z}\mathbf{h}
    +\sqrt{\varrho_\text{r}}\boldsymbol{\nu}+\mathbf{w},
\end{align}
where $\mathbf{y}=[y_1,\cdots,y_T]^\mathrm{T}$, $\mathbf{Z}=[\mathbf{z}_1^\mathrm{T},\cdots,
\mathbf{z}_T^\mathrm{T}]^\mathrm{T}$, $\boldsymbol{\nu}=[\mathbf{z}_1\mathbf{v}_1,
\cdots,\mathbf{z}_T\mathbf{v}_T]^\mathrm{T}$ with $\mathbf{v}_t=[a_{t,1}g_{\text{r},1}^{*},\cdots,
a_{t,N}g_{\text{r},N}^{*}]$. Since $\mathbf{h}$ is an $N\times1$ vector and $\mathbf{Z}$ is an $T\times N$ matrix, to ensure that $\mathbf{h}$ can be uniquely estimated, the number of RIS training patterns must satisfy that $T\geq N$. A $T\times T$ discrete Fourier transform (DFT) matrix or Hadamard matrix may then be employed for the design of RIS training patterns \cite{zhou2020joint}. Specifically, $[\text{e}^{\jmath\overline{\theta}_{t,1}},\cdots,
\text{e}^{\jmath\overline{\theta}_{t,N}}]$ is designed as the first $N$ elements in the $t$th row of the $T\times T$ discrete Fourier transform (DFT) matrix or Hadamard matrix. Our purpose is to estimating $\mathbf{h}$ based on the received signal $\mathbf{y}$ contaminated by the thermal noise of the active reflection amplifier and the additive white Gaussian noise at the receiver, which can be estimated using different channel estimation methods such as the LMMSE estimator as discussed in follows.

\begin{corollary}\label{corollary_1}
The LMMSE estimate of $\mathbf{h}$ is based on the observation $\mathbf{y}$, formulated as:
\begin{align}\label{Channel_Estimation_3}
    \hat{\mathbf{h}}=\mathbb{E}[\mathbf{h}]+\mathbf{C}_{\mathbf{h}\mathbf{y}}
    \mathbf{C}_{\mathbf{y}\mathbf{y}}^{-1}(\mathbf{y}-\mathbb{E}[\mathbf{y}]),
\end{align}
where $\mathbb{E}[\mathbf{h}]=[\overline{g}_{\text{t},1}\overline{g}_{\text{r},1}^{*},
\cdots,\overline{g}_{\text{t},N}
\overline{g}_{\text{r},N}^{*}]^{\text{T}}$, $\mathbb{E}[\mathbf{y}]=\sqrt{\rho\varrho_\text{t}\varrho_\text{r}}\xi
\overline{\mathbf{Z}}\mathbb{E}[\mathbf{h}]$, $\mathbf{C}_{\mathbf{h}\mathbf{y}}$ and $\mathbf{C}_{\mathbf{y}\mathbf{y}}$ are given as follows
\begin{align}\label{Channel_Estimation_6}
    \mathbf{C}_{\mathbf{h}\mathbf{y}}=\sqrt{\rho\varrho_\text{t}\varrho_\text{r}}\eta\xi
    \overline{\mathbf{Z}}^{\text{H}},
\end{align}
\begin{align}\label{Channel_Estimation_7}
    \mathbf{C}_{\mathbf{y}\mathbf{y}}=\rho\varrho_\text{t}\varrho_\text{r}
    (\xi^2\eta\overline{\mathbf{Z}}\overline{\mathbf{Z}}^{\text{H}}+(1-\xi^2)
    \ddot{\beta}\mathbf{I}_T)+\varrho_\text{r}\sigma_a^2\ddot{\beta}\mathbf{I}_T
    +\sigma_w^2\mathbf{I}_T,
\end{align}
where we have $\ddot{\beta}=\sum_{n=1}^{N}\beta_n^2$, $\eta=\frac{1+\kappa_t+\kappa_r}{1+\kappa_t+\kappa_r+\kappa_t\kappa_r}$, $\xi=\frac{I_1(\varsigma_\text{p})}{I_0(\kappa_\text{p})}$ when the RIS phase noise obeys $\mathcal{VM}(0,\varsigma_p)$, and $\xi=\frac{\sin(\iota_\text{p})}{\iota_\text{p}}$ when the RIS phase noise follows $\mathcal{U}(-\iota_\text{p},\iota_\text{p})$, with $I_k(\cdot)$ representing the modified Bessel functions of the first kind of order $k$. Furthermore, we have $\overline{\mathbf{Z}}=[\overline{\mathbf{z}}_1^{\text{T}},\overline{\mathbf{z}}_2^{\text{T}},
\cdots,\overline{\mathbf{z}}_T^{\text{T}}]^{\text{T}}$, where $\overline{\mathbf{z}}_t=[\beta_1\text{e}^{\jmath\overline{\theta}_{t,1}},
\beta_2\text{e}^{\jmath\overline{\theta}_{t,2}},\cdots,
\beta_N\text{e}^{\jmath\overline{\theta}_{t,N}}]$.
\end{corollary}
\begin{IEEEproof}
    See Appendix \ref{Appendix_A}.
\end{IEEEproof}

\begin{corollary}\label{corollary_2}
The covariance matrix of the estimated channel $\hat{\mathbf{h}}$ is
\begin{align}\label{Channel_Estimation_8}
    \mathbf{C}_{\hat{\mathbf{h}}\hat{\mathbf{h}}}=\text{diag}\{\varepsilon_1,
    \varepsilon_2,\cdots,\varepsilon_N\},
\end{align}
where $\varepsilon_n=\frac{T\rho\varrho_\text{t}\varrho_\text{r}\xi^2\beta_n^2\eta^2}
{T\rho\varrho_\text{t}\varrho_\text{r}\xi^2\beta_n^2\eta+\rho\varrho_\text{t}
\varrho_\text{r}(1-\xi^2)\ddot{\beta}+\varrho_\text{r}\ddot{\beta}
\sigma_a^2+\sigma_w^2}$.
The estimation error is $\tilde{\mathbf{h}}=\mathbf{h}-\hat{\mathbf{h}}$, and its covariance matrix is
\begin{align}\label{Channel_Estimation_10}
    \mathbf{C}_{\tilde{\mathbf{h}}\tilde{\mathbf{h}}}=\text{diag}
    \{\eta-\varepsilon_1,\eta-\varepsilon_2,\cdots,\eta-\varepsilon_N\}.
\end{align}
\end{corollary}
\begin{IEEEproof}
    See Appendix \ref{Appendix_B}.
\end{IEEEproof}

According to (\ref{Channel_Estimation_10}), the normalized mean square error (N-MSE), denoted as $\epsilon$, is given by
\begin{align}\label{Channel_Estimation_11}
    \epsilon=\frac{\mathbb{E}[\|\tilde{\mathbf{h}}\|^2]}{\mathbb{E}
    [\|\mathbf{h}\|^2]}=\frac{1}{N}\text{Tr}[\mathbf{C}_{\tilde{\mathbf{h}}
    \tilde{\mathbf{h}}}]=\eta-\frac{1}{N}\sum\nolimits_{n=1}^{N}\varepsilon_n.
\end{align}
From (\ref{Channel_Estimation_11}), we can see that when the transmit power obeys $\rho\rightarrow\infty$, the N-MSE becomes $\epsilon=\frac{1}{N}\sum_{n=1}^{N}\frac{(1-\xi^2)\ddot{\beta}\eta}
{T\xi^2\beta_n^2\eta+(1-\xi^2)\ddot{\beta}}$, which indicates that the N-MSE tends to a non-zero floor, instead of 0. In the high power region this is due to the effect of RIS phase noise. For passive RIS, associated with $\beta_1=\beta_2=\cdots=\beta_N=1$, the N-MSE obeys $\epsilon=\eta-\frac{T\rho\varrho_\text{t}\varrho_\text{r}\xi^2\eta^2}
{T\rho\varrho_\text{t}\varrho_\text{r}\xi^2\eta+\rho\varrho_\text{t}\varrho_\text{r}
(1-\xi^2)N+\varrho_\text{r}N\sigma_a^2+\sigma_w^2}$.

\section{Performance Analysis}\label{Performance_Analysis}
Once $\hat{\mathbf{h}}$ becomes known, the RIS phase shift during the data transmission can be designed as $\overline{\theta}_n=-\angle\hat{h}_n$ for coherently combining the signals reflected by all RIS elements \cite{basar2019wireless}. In data transmission, when the desired RIS configuration vector is $\overline{\mathbf{z}}=[\beta_1\text{e}^{\jmath\overline{\theta}_1},\cdots,
\beta_N\text{e}^{\jmath\overline{\theta}_N}]$ and the RIS phase noise are $\widetilde{\theta}_1,
\cdots,\widetilde{\theta}_N$, the practical RIS configuration vector is $\mathbf{z}=[\beta_1\text{e}^{\jmath\theta_1},\cdots,\beta_N\text{e}^{\jmath\theta_N}]
=[\beta_1\text{e}^{\jmath(\overline{\theta}_1+\widetilde{\theta}_1)},\cdots,
\beta_N\text{e}^{\jmath(\overline{\theta}_N+\widetilde{\theta}_N)}]$. Thus, the received symbol is given by
\begin{align}\label{Performance_Analysis_1}
    \notag y=&\underbrace{\sqrt{\rho\varrho_\text{t}\varrho_\text{r}}\overline{\mathbf{z}}
    \hat{\mathbf{h}}s}_{\text{Desired signal over estimated channel}}+
    \underbrace{\sqrt{\rho\varrho_\text{t}\varrho_\text{r}}(\mathbf{z}-\overline{\mathbf{z}}
    )\hat{\mathbf{h}}s}_{\text{RIS phase noise over estimated channel}}\\
    \notag&+\underbrace{\sqrt{\rho\varrho_\text{t}\varrho_\text{r}}\mathbf{z}
    \tilde{\mathbf{h}}s}_{\text{Signal over unknown channel}}+
    \underbrace{\sqrt{\varrho_\text{r}}\mathbf{z}\mathbf{v}}_{\text{Thermal noise from RIS reflection amplifiers}}\\
    &+\underbrace{w}_{\text{Receiver additive noise}},
\end{align}
where $w\sim\mathcal{CN}(0,\sigma_w^2)$, $\mathbf{v}=[a_1g_{\text{r},1}^{*},\cdots,a_Ng_{\text{r},N}^{*}]^{\text{T}}$ is the cascaded channel of $\mathbf{g}_{\text{t}}$ and the thermal noise of the active reflection amplifier $\mathbf{a}=[a_1,\cdots,a_N]^\mathrm{T}$ with $\mathbf{a}\sim\mathcal{CN}(\mathbf{0}_N,\sigma_a^2\mathbf{I}_N)$. In (\ref{Performance_Analysis_1}), the first item represents the desired signal, while the other items can be viewed as equivalent noise, since only their statistical characteristics are known. Referring to \cite{long2021active,zhi2022active}, the average energy cost at the active RIS is $\mathbb{E}[\rho\varrho_\mathrm{t}|\mathbf{z}\mathbf{g}_\mathrm{t}|^2+
|\mathbf{z}\mathbf{a}|^2]+N(P_s+P_d)=(\rho\varrho_\mathrm{t}+\sigma_a^2)
\sum\nolimits_{n=1}^{N}\beta_n^2+N(P_\mathrm{s}+P_\mathrm{d})$, where $P_\mathrm{s}$ represents the power consumed by the phase shift switch and control circuit for each reflecting element and $P_\mathrm{d}$ is the direct current biasing power used by each amplifier.

The power of the desired signal received over the estimated channel is
\begin{align}\label{Performance_Analysis_2}
    X_\text{s}=\rho\varrho_\text{t}\varrho_\text{r}|\overline{\mathbf{z}}
    \hat{\mathbf{h}}|^2\overset{(a)}{=}\rho\varrho_\text{t}\varrho_\text{r}\xi^2
    \big(\sum\nolimits_{n=1}^{N}\beta_n|\hat{h}_n|\big)^2,
\end{align}
where (a) is based on $\overline{\theta}_n=-\angle\hat{h}_n$ and $\mathbb{E}[\text{e}^{\jmath\theta_n}]=\xi\text{e}^{\jmath\overline{\theta}_n}$.

Since $\mathbb{E}[\hat{\mathbf{h}}\tilde{\mathbf{h}}^{\text{H}}]=\mathbf{O}_N$, and $a_n$ and $w$ are independent of $\hat{\mathbf{h}}$ and $\tilde{\mathbf{h}}$, the power of the equivalent noise, denoted by $X_\text{n}$, is
\begin{align}\label{Performance_Analysis_3}
    X_\text{n}=X_\text{n,p}+X_\text{n,e}+X_\text{n,a}+X_\text{n,w},
\end{align}
where $X_\text{n,p}$, $X_\text{n,e}$, $X_\text{n,a}$ and $X_\text{n,w}$ represent the power of RIS phase noise contribution received over the estimated channel, the desired signal over an unknown channel, the noise from the RIS reflection amplifiers and the receiver additive noise, respectively. Firstly, the variance of the RIS phase noise over the estimated channel is formulated as:
\begin{align}\label{Performance_Analysis_4}
    \notag X_\text{n,p}=&\rho\varrho_\text{t}\varrho_\text{r}\mathbb{E}
    [|(\mathbf{z}-\overline{\mathbf{z}})\hat{\mathbf{h}}|^2]\\
    \notag=&\rho\varrho_\text{t}\varrho_\text{r}\sum\nolimits_{n_1=1}^{N}\sum\nolimits_{n_2=1}^{N}
    \mathbb{E}[\beta_{n_1}(\text{e}^{\jmath(\bar{\theta}_{n_1}+\tilde{\theta}_{n_1})}
    -\xi\text{e}^{\jmath\overline{\theta}_{n_1}})\cdot\\
    \notag&\hat{h}_{n_1}\hat{h}_{n_2}^{*}(\text{e}^{-\jmath(\overline{\theta}_{n_2}
    +\tilde{\theta}_{n_2})}-\xi\text{e}^{-\jmath\overline{\theta}_{n_2}})\beta_{n_2}]\\
    \notag\overset{(a)}{=}&\rho\varrho_\text{t}\varrho_\text{r}
    \sum\nolimits_{n_1=1}^{N}\sum\nolimits_{n_2=1}^{N}
    \beta_{n_1}\beta_{n_2}|\hat{h}_{n_1}||\hat{h}_{n_2}|\cdot\\
    \notag&\mathbb{E}[(\text{e}^{\jmath\widetilde{\theta}_{n_1}}-\xi)
    (\text{e}^{-\jmath\tilde{\theta}_{n_2}}-\xi)]\\
    \overset{(b)}{=}&\rho\varrho_\text{t}\varrho_\text{r}(1-\xi^2)\sum\nolimits_{n=1}^{N}
    \beta_{n}^2\varepsilon_n,
\end{align}
where (a) is based on $|\hat{h}_n|=\text{e}^{\jmath\overline{\theta}_n}\hat{h}_n$, and (b) is based on the fact that $\mathbb{E}[(\text{e}^{\jmath\widetilde{\theta}_{n_1}}-\xi)
(\text{e}^{-\jmath\tilde{\theta}_{n_2}}-\xi)]$ equals to $1-\xi^2$ when $n_1=n_2$ and equals to 0 when $n_1\neq n_2$. Then, the power of the desired signal over an unknown channel is
\begin{align}\label{Performance_Analysis_5}
    \notag{X}_\text{n,e}&=\rho\varrho_\text{t}\varrho_\text{r}\mathbb{E}[|\sum\nolimits_{n=1}^{N}
    \beta_n\text{e}^{\jmath\theta_n}\tilde{h}_n|^2]\\
    &\overset{(a)}{=}
    \rho\varrho_\text{t}\varrho_\text{r}\sum\nolimits_{n=1}^{N}\beta_n^2(\eta-\varepsilon_n),
\end{align}
where (a) is derived according to (\ref{Channel_Estimation_10}).
Next, the power of the noise emanating from the RIS reflection amplifiers is:
\begin{align}\label{Performance_Analysis_6}
    X_\text{n,a}=\varrho_\text{r}\mathbb{E}\Big[\big|\sum\nolimits_{n=1}^{N}
    \beta_n\text{e}^{\jmath\theta_n}a_ng_{\text{r},n}^{*}\big|^2\Big]\overset{(a)}{=}
    \varrho_\text{r}\ddot{\beta}\sigma_a^2,
\end{align}
where (a) is based on the independence of $a_n$ ($n=1,\cdots,N$). Finally, the power of noise in the receiver is
\begin{align}\label{Performance_Analysis_7}
    X_\text{n,w}=\sigma_w^2.
\end{align}
Upon substituting (\ref{Performance_Analysis_4}), (\ref{Performance_Analysis_5}), (\ref{Performance_Analysis_6}) and (\ref{Performance_Analysis_7}) into (\ref{Performance_Analysis_3}), we can express the power of the equivalent noise as
\begin{align}\label{Performance_Analysis_8}
    X_\text{n}=\rho\varrho_\text{t}\varrho_\text{r}\sum\nolimits_{n=1}^{N}\beta_n^2
    (\eta-\xi^2\varepsilon_n)+\varrho_\text{r}\ddot{\beta}\sigma_a^2+\sigma_w^2.
\end{align}

According to (\ref{Performance_Analysis_2}) and (\ref{Performance_Analysis_8}), we can express the instantaneous received SINR, denoted by $\gamma$, as
\begin{align}\label{Performance_Analysis_9}
    \gamma=\frac{X_\text{s}}{X_\text{n}}=\gamma_0\big(\sum\nolimits_{n=1}^{N}\beta_n|\hat{h}_n|\big)^2,
\end{align}
where we have $\gamma_0=\frac{\rho\varrho_\text{t}\varrho_\text{r}\xi^2}{\rho\varrho_\text{t}
\varrho_\text{r}\sum_{n=1}^{N}\beta_n^2(\eta-\xi^2\varepsilon_n)+\varrho_\text{r}
\ddot{\beta}\sigma_a^2+\sigma_w^2}$. To derive the closed-form distribution of the instantaneous received SINR $\gamma$, we employ the moment-matching method to approximate it as follows.
\begin{corollary}\label{corollary_3}
Firstly, the first moment and the second moment of $\gamma$ can be approximated, derived as
\begin{align}\label{Performance_Analysis_10}
    \notag\mathbb{E}[\gamma]=&\gamma_0\Big[\sum\nolimits_{n=1}^{N}\beta_n^2
    (1-\eta+\varepsilon_n)
    +\frac{\pi}{4}\sum\nolimits_{(n_1,n_2)=1}^{N}\beta_{n_1}\beta_{n_2}\\
    &\sqrt{\varepsilon_{n_1}\varepsilon_{n_2}}L_{\frac{1}{2}}
    \Big(\frac{\eta-1}{\varepsilon_{n_1}}\Big)L_{\frac{1}{2}}
    \Big(\frac{\eta-1}{\varepsilon_{n_2}}\Big)\Big],
\end{align}
\begin{align}\label{Performance_Analysis_11}
    \notag\mathbb{E}[\gamma^2]=&\gamma_0^2\Big[\sum\nolimits_{n=1}^{N}\beta_n^4
    (2\varepsilon_n^2+4\varepsilon_n(1-\eta)+(1-\eta)^2)\\
    \notag&+\frac{3\pi}{2}\sum_{(n_1,n_2)=1}^{N}\beta_{n_1}^3\beta_{n_2}
    \varepsilon_{n_1}^{\frac{3}{2}}\varepsilon_{n_2}^{\frac{1}{2}}L_{\frac{3}{2}}
    \Big(\frac{\eta-1}{\varepsilon_{n_1}}\Big)L_{\frac{1}{2}}
    \Big(\frac{\eta-1}{\varepsilon_{n_2}}\Big)\\
    \notag&+6\sum_{(n_1,n_2)=1}^{N}\beta_{n_1}^2\beta_{n_2}^2
    (\varepsilon_{n_1}+1-\eta)(\varepsilon_{n_2}+1-\eta)\\
    \notag&+\frac{3}{2}\pi^2\sum_{(n_1,n_2,n_3,n_4)=1}^{N}\beta_{n_1}
    \beta_{n_2}\beta_{n_3}\beta_{n_4}\sqrt{\varepsilon_{n_1}\varepsilon_{n_2}\varepsilon_{n_3}
    \varepsilon_{n_4}}\\
    \notag&L_{\frac{1}{2}}\Big(\frac{\eta-1}{\varepsilon_{n_1}}\Big)L_{\frac{1}{2}}
    \Big(\frac{\eta-1}{\varepsilon_{n_2}}\Big)L_{\frac{1}{2}}\Big(\frac{\eta-1}
    {\varepsilon_{n_3}}\Big)L_{\frac{1}{2}}\Big(\frac{\eta-1}{\varepsilon_{n_4}}\Big)\\
    \notag&+3\pi\sum_{(n_1,n_2,n_3)=1}^{N}\beta_{n_1}^2\beta_{n_2}\beta_{n_3}
    (\varepsilon_{n_1}+1-\eta)\sqrt{\varepsilon_{n_2}\varepsilon_{n_3}}\\
    &L_{\frac{1}{2}}\Big(\frac{\eta-1}{\varepsilon_{n_2}}\Big)L_{\frac{1}{2}}
    \Big(\frac{\eta-1}{\varepsilon_{n_3}}\Big)\Big],
\end{align}
where $L_{p}(\cdot)$ represents the Laguerre polynomial of degree $p$.
\end{corollary}
\begin{IEEEproof}
    See Appendix \ref{Appendix_C}.
\end{IEEEproof}

Based on \cite{van2020coverage}, the received SINR $\gamma$ can be approximated by the Gamma distribution, denoted as $\gamma\sim\mathcal{GM}(\upsilon,\vartheta)$, where the shape parameter $\upsilon$ and the scale parameter $\vartheta$ are given by
\begin{align}\label{Performance_Analysis_13}
    \upsilon=\frac{\mathbb{E}[\gamma]^2}{\mathbb{V}[\gamma]},\ \vartheta=\frac{\mathbb{V}[\gamma]}{\mathbb{E}[\gamma]},
\end{align}
where we have $\mathbb{V}[\gamma]=\mathbb{E}[\gamma^2]-\mathbb{E}[\gamma]^2$. Based on the distribution of the instantaneous received SINR, we can calculate the outage probability and ergodic channel capacity as follows.

\subsection{Outage Probability}
The outage probability $P_\text{o}$ is routinely used for characterizing the reliability of communication links, which is defined as the probability of the received SINR $\gamma$ being lower than a given threshold $\omega_\text{th}$, i.e. $P_\text{o}=\Pr(\gamma<\omega_\text{th})$. Based on (\ref{Performance_Analysis_11}), we can express the outage probability as
\begin{align}\label{Outage_Probability_1}
    P_\text{o}=\Pr(\gamma<\omega_\text{th})=F_{\gamma}(\omega_\text{th})
    \overset{(a)}=\frac{1}{\Gamma(\upsilon)}\Gamma^{\text{li}}
    (\upsilon,\frac{\omega_\text{th}}{\vartheta}),
\end{align}
where $\upsilon$ and $\vartheta$ are given in (\ref{Performance_Analysis_13}), (a) is based on $\gamma\sim\mathcal{GM}(\upsilon,\vartheta)$ and the PDF of a Gamma distribution, $\Gamma(\upsilon)$ is the Gamma function of $\upsilon$, and $\Gamma^{\text{li}}(\upsilon,\frac{\omega_\text{th}}{\vartheta})$ represents the lower incomplete Gamma function of $\frac{\omega_\text{th}}{\vartheta}$ with integration range $(0,\upsilon)$.

\subsection{Ergodic Channel Capacity}
The ergodic channel capacity, denoted as $R$, is the average of the instantaneous channel capacity, given by
\begin{align}\label{Ergodic_Channel_Capacity_1}
    R=\mathbb{E}[\log_2(1+\gamma)]=\log_2\mathrm{e}\cdot\mathbb{E}[\ln(\gamma')],
\end{align}
where $\gamma'=1+\gamma$. It is then plausible that $\mathbb{E}[\gamma']=1+\mathbb{E}[\gamma]$ and $\mathbb{V}[\gamma']=\mathbb{V}[\gamma]$. Thus, $\gamma'$ is approximately follows the Gamma distribution, i.e. $\gamma'\sim\mathcal{GM}(\upsilon',\vartheta')$, with the shape parameter and scale parameter given by:
\begin{align}\label{Ergodic_Channel_Capacity_2}
    \upsilon'=\frac{(\mathbb{E}[\gamma]+1)^2}{\mathbb{V}[\gamma]},\ \vartheta'=\frac{\mathbb{V}[\gamma]}{\mathbb{E}[\gamma]+1}.
\end{align}
Since the logarithmic expectation of the variable $X$ obeying the Gamma distribution, given by $\mathcal{GM}(\upsilon',\vartheta')=\log_2\mathrm{e}\cdot\psi(\upsilon')
+\log_2\vartheta'$, where $\psi(\cdot)$ represents the digamma function, we can express the ergodic channel capacity as
\begin{align}\label{Ergodic_Channel_Capacity_3}
    R=\log_2\mathrm{e}\cdot\psi\Big(\frac{(\mathbb{E}[\gamma]+1)^2}{\mathbb{V}
    [\gamma]}\Big)+\log_2\frac{\mathbb{V}[\gamma]}{\mathbb{E}[\gamma]+1},
\end{align}
where $\mathbb{V}[\gamma]=\mathbb{E}[\gamma^2]-\mathbb{E}[\gamma]^2$, and $\mathbb{E}[\gamma]$ and $\mathbb{E}[\gamma^2]$ are given in (\ref{Performance_Analysis_10}) and (\ref{Performance_Analysis_11}), respectively.

If we assume the RIS amplitude gain of all reflecting elements to be the same, i.e. $\beta_1=\beta_2=\cdots=\beta_N=\beta$, then we have $\varepsilon_1=\varepsilon_2=\cdots=\varepsilon_N
=\frac{T\xi^2\eta^2}{T\xi^2\eta+N(1-\xi^2)}=\varepsilon$. Hence, we can get the following asymptotic ergodic channel capacity scaling law with respect to the number of RIS elements $N$.
\begin{corollary}\label{corollary_4}
When $N\rightarrow\infty$ and $\rho\rightarrow\infty$, the ergodic channel capacity follows
\begin{align}\label{Ergodic_Channel_Capacity_4}
    R\rightarrow\log_2\Big(1+\frac{\pi}{4}\frac{\xi^2\varepsilon}{\eta-\xi^2\varepsilon}
    L_{\frac{1}{2}}^2\Big(\frac{\eta-1}{\varepsilon}\Big)N\Big).
\end{align}
Specifically, in the Rayleigh fading channel, $\eta=1$ and the ergodic channel capacity obeys:
\begin{align}\label{Ergodic_Channel_Capacity_5}
    R\rightarrow\log_2\Big(1+\frac{\pi}{4}\frac{\xi^2\varepsilon}{1-\xi^2\varepsilon}N\Big).
\end{align}
\end{corollary}
\begin{IEEEproof}
    See Appendix \ref{Appendix_D}.
\end{IEEEproof}

\section{Numerical and Simulation Results}\label{Numerical_and_Simulation_Results}
\begin{figure}[!t]
    \centering
    \includegraphics[width=2.9in]{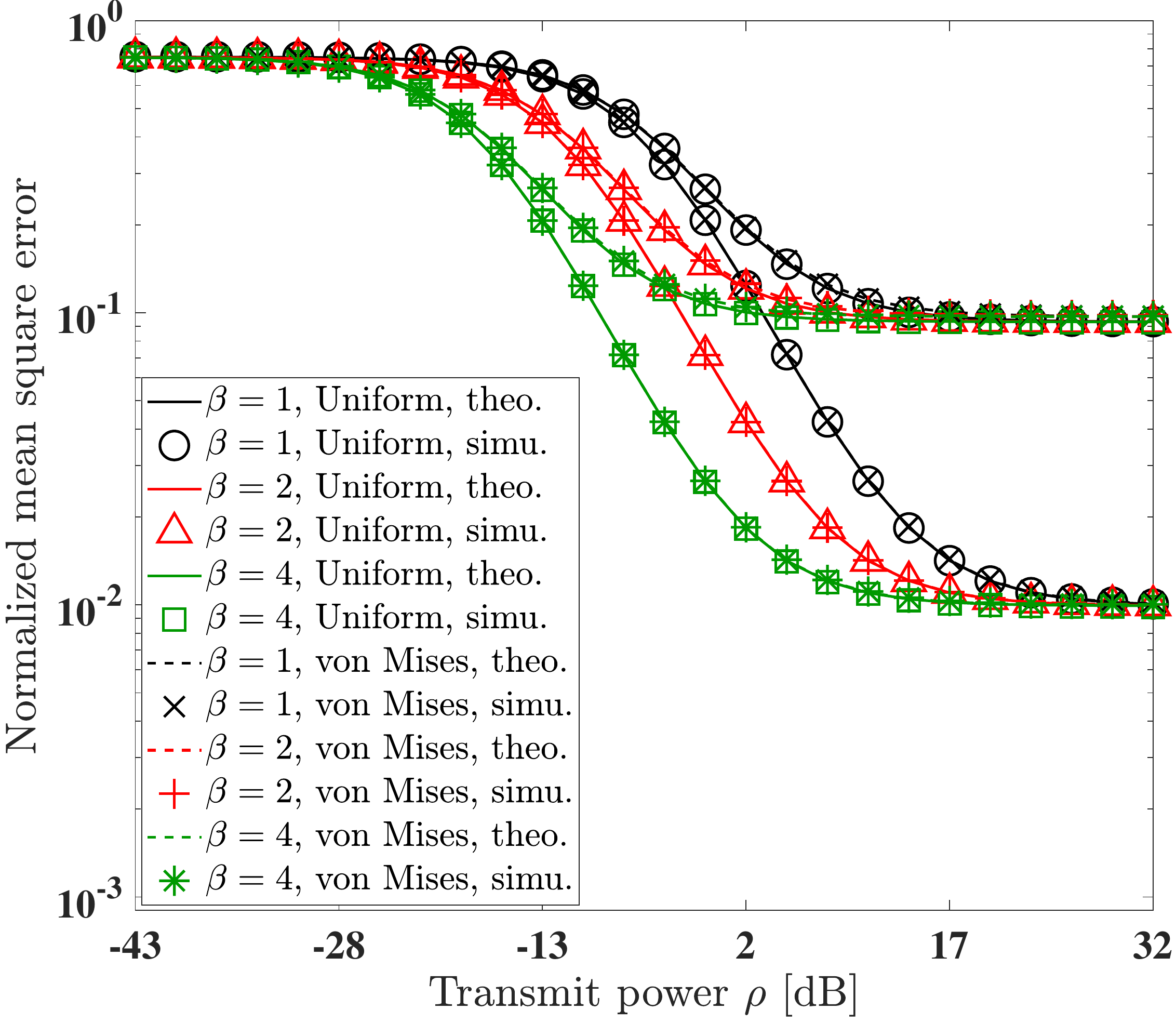}
    \caption{Comparison of the normalized mean square error performance versus the transmit power $\rho$.}\label{Fig_simu_MSE_active_RIS}
\end{figure}

\begin{figure}[!t]
    \centering
    \includegraphics[width=2.9in]{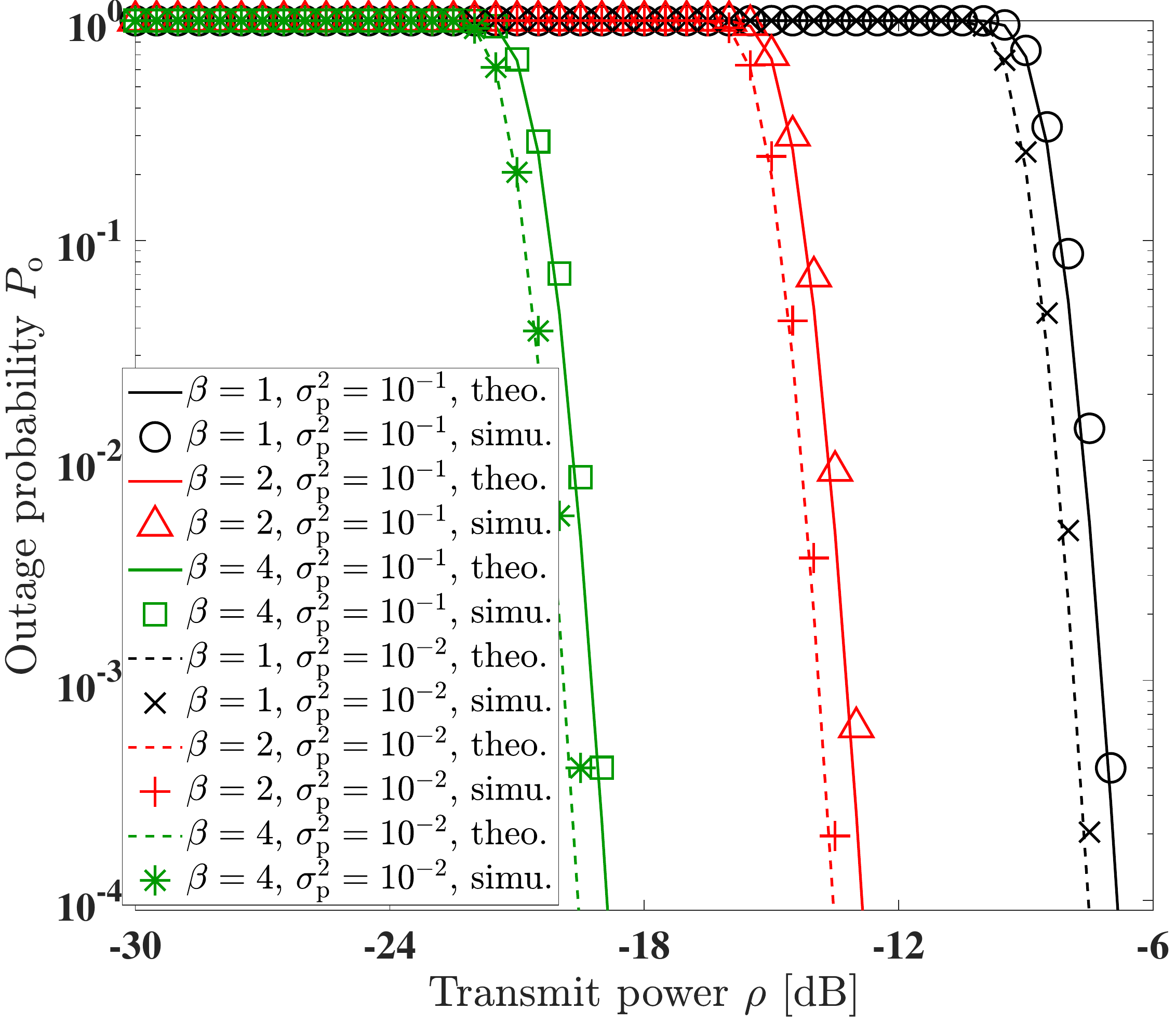}
    \caption{Comparison of the outage probability ($P_\text{o}$) versus the transmit power $\rho$.}\label{Fig_simu_P_out_active_RIS}
\end{figure}

\begin{figure}[!t]
    \centering
    \includegraphics[width=2.9in]{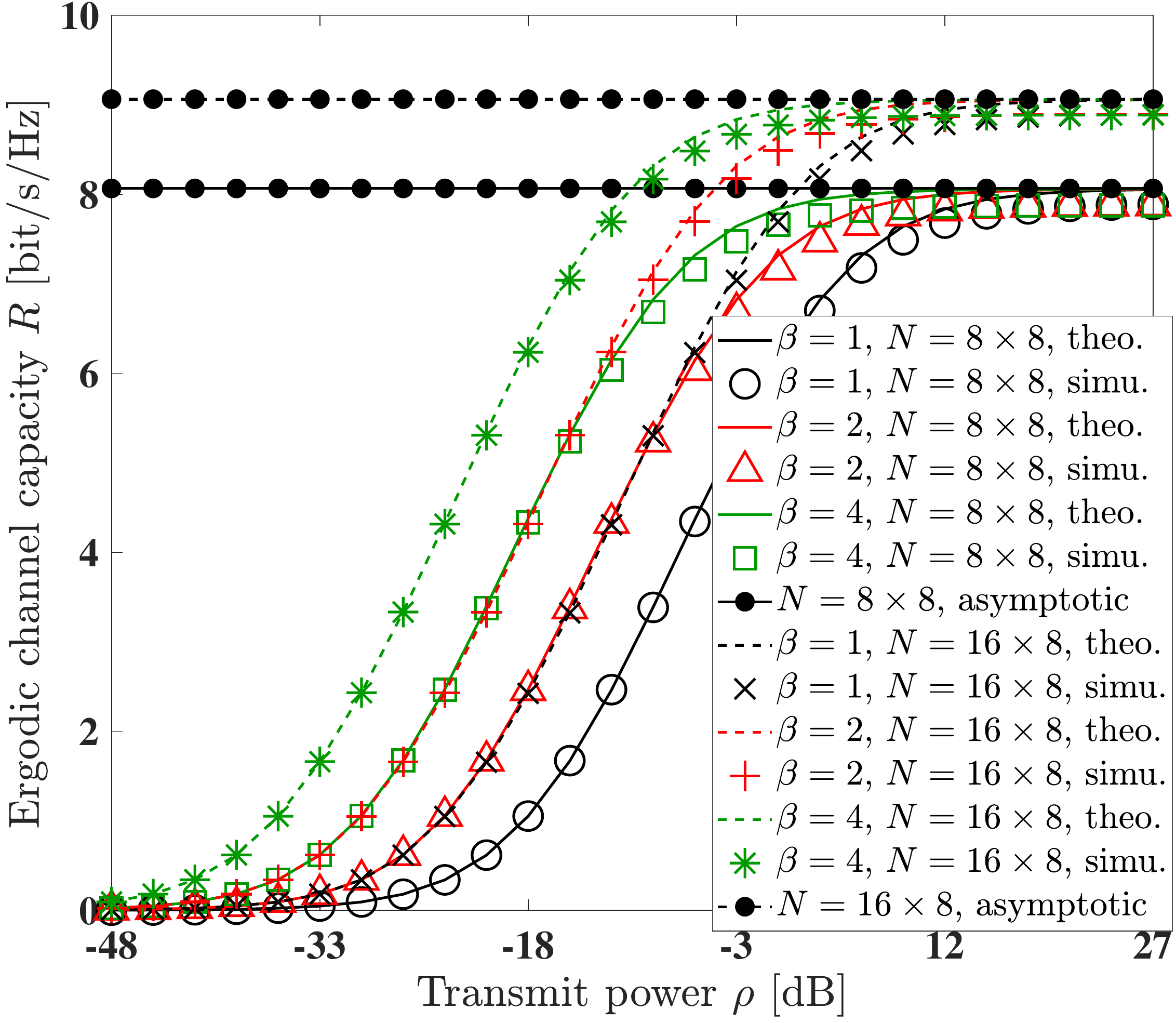}
    \caption{Comparison of the ergodic channel capacity ($R$) versus the transmit power $\rho$.}\label{Fig_simu_SE_active_RIS}
\end{figure}

In this section, the theoretical and simulation results of the N-MSE at the stage of channel estimation, and the outage probability (OP) as well as the ergodic channel capacity at the stage of data transmission, are presented. The transmitter, the RIS, and the receiver are located at the Cartesian coordinates of (-100m, 0m, 15m), (0m, 50m, 15m), (15m, 45m, 15m), respectively. Unless otherwise specified, the simulation parameters are: $\kappa_\text{t}=\kappa_\text{r}=0\text{dB}$, $\alpha_\text{t}=\alpha_\text{r}=2.2$, $\mathrm{C}_0=-30\text{dB}$, $\sigma_a^2=-90\text{dBm}$, $\sigma_w^2=-90\text{dBm}$, $N=8\times8$, and $\delta_x=\delta_y=\frac{\lambda}{2}$. The number of RIS training patterns is $T=N$, which are designed based on the DFT matrix.

The theoretical analysis (theo.) and the simulation results (simu.) of the normalized mean square error performance of the LMMSE estimator is shown in Fig. \ref{Fig_simu_MSE_active_RIS}, with the RIS phase shift noise characterized by $\sigma_\text{p}^2=10^{-2},10^{-1}$, and the RIS amplitude gain of all reflecting elements being identical, i.e. $\beta=\beta_1=\beta_2=\cdots=\beta_N$, and $\beta=1,2,4$ respectively. Fig. \ref{Fig_simu_MSE_active_RIS} shows that upon increasing the transmit power $\rho$, the N-MSE tends to a constant value when the RIS phase noise power obeys $\sigma_\text{p}^2>0$. This is because the RIS phase noise escalates upon increasing the transmit power. Furthermore, similar N-MSE performance is achieved, when the RIS phase noise follows the uniform distribution and the von Mises distribution having the same phase noise power.

To evaluate further, the OP and ergodic channel capacity are shown in Fig. \ref{Fig_simu_P_out_active_RIS} and Fig. \ref{Fig_simu_SE_active_RIS}, respectively. Firstly, in Fig. \ref{Fig_simu_P_out_active_RIS}, the RIS phase noise follows the uniform distribution with $\sigma_\text{p}^2=10^{-1}$ and $\sigma_\text{p}^2=10^{-2}$, and the threshold SINR is $\omega_\text{th}=10\text{dB}$. Observe that the OP is reduced extremely rapidly, since using a large number of RIS elements for coherently combining the received signals hardens the channel and combats the channel fading. Secondly, in Fig. \ref{Fig_simu_SE_active_RIS}, the RIS phase noise follows the uniform distribution having $\sigma_\text{p}^2=10^{-1}$. Observe in Fig. \ref{Fig_simu_SE_active_RIS} that a slight gap exists between the simulation results and the theoretical analysis, which arises from the fact that we approximate the received SNR using the Gamma distribution. In contrast to the perfect CSI scenario, the spectral efficiency saturates upon increasing of the transmit power due to both the channel estimation error and the RIS phase shift noise. This can be explained using (\ref{Performance_Analysis_1}), where the contribution of RIS phase noise over the estimated channel and that of the signal over the unknown channel increase with the transmit power. When doubling the number of RIS elements we have about 1 bit/s/Hz spectral efficiency improvement in the high transmit power region. We can now readily see that the ergodic channel capacity asymptotically tends to $\log_2\Big(1+\frac{\pi}{4}\frac{\xi^2\varepsilon}{\eta-\xi^2\varepsilon}
L_{\frac{1}{2}}^2\Big(\frac{\eta-1}{\varepsilon}\Big)N\Big)$ in the high transmit power region. Furthermore, as seen from Fig. \ref{Fig_simu_MSE_active_RIS} to Fig. \ref{Fig_simu_SE_active_RIS}, about $6\text{dB}$ performance enhancement can be attained by doubling the RIS amplitude gain $\beta$ at the cost of additional power dissipation. This reveals that in the low transmit power region, harnessing more RIS elements and increasing the RIS amplifier power can improve the ergodic channel capacity. By contrast, in the high transmit power region, only employment of more RIS elements can effectively improve the ergodic channel capacity.

\section{Conclusions}\label{Conclusion}
We formulated the LMMSE channel estimation prototype of active RIS-aided wireless communication in the face of RIS phase shift noise and derived the corresponding theoretical normalized mean square error. Then, the theoretical analysis of the OP and ergodic channel capacity was provided based on the moment-match method for approximating the distribution of the received SINR, while considering the effects of channel estimation error, the thermal noise of the RIS amplifiers and the RIS phase noise.

\appendices
\section{}\label{Appendix_A}
Eq (\ref{Channel_Estimation_3}) can be derived based on the LMMSE criterion, while $\mathbb{E}[\mathbf{h}]$ can be derived according to (\ref{Channel_Model_2}), (\ref{Channel_Model_3}) and (\ref{Channel_Model_4}).

Since the RIS phase noise $\tilde{\theta}_{t,n}$ has i.i.d. for $t=1,\cdots,T$ and $n=1,\cdots,N$, we can omit the subscripts in $\tilde{\theta}_{t,n}$ for simplicity. Given that the RIS phase noise is also independent of $\overline{\mathbf{Z}}$, we can write the mean of $\mathbf{Z}$ as $\mathbb{E}[\mathbf{Z}]=\mathbb{E}[\text{e}^{\jmath\tilde{\theta}}]
\overline{\mathbf{Z}}$. Firstly, when the RIS phase noise obeys $\tilde{\theta}\sim\mathcal{VM}(0,\varsigma_\text{p})$, upon referring to \cite{hillen2017moments}, we have $\mathbb{E}[\text{e}^{\jmath\tilde{\theta}}]=\frac{I_{1}(\varsigma_\text{p})}
{I_{0}(\varsigma_\text{p})}$. Thus, we can infer that $\mathbb{E}[\mathbf{Z}]=\xi\overline{\mathbf{Z}}$ with $\xi=\frac{I_{1}(\varsigma_\text{p})}{I_{0}(\varsigma_\text{p})}$ if $\tilde{\theta}\sim\mathcal{VM}(0,\varsigma_\text{p})$. Secondly, if the RIS phase noise obeys $\tilde{\theta}\sim\mathcal{UF}(-\iota_\mathrm{p},\iota_\mathrm{p})$, the $i$th-order moment of $\tilde{\theta}$, namely $\mathbb{E}[{\tilde{\theta}}^i]$, is equal to 0 when $i$ is odd and equal to $\frac{1}{i+1}\iota_\mathrm{p}^i$ when $i$ is even. Thus, we can show that
\begin{align}\label{Appendix_A_1}
    \mathbb{E}[e^{j\tilde{\theta}}]&=\sum_{i=0}^{\infty}\frac{(-1)^i}{(2i)!}\mathbb{E}
    [\tilde{\theta}^{2i}]=\sum_{i=0}^{\infty}\frac{(-1)^i\iota_\mathrm{p}^{2i}}{(2i+1)!}
    =\frac{\sin(\iota_\mathrm{p})}{\iota_\mathrm{p}}.
\end{align}
Hence, we can show that $\mathbb{E}[\mathbf{Z}]=\xi\overline{\mathbf{Z}}$ with $\xi=\frac{\sin(\iota_\mathrm{p})}{\iota_\mathrm{p}}$, when $\tilde{\theta}\sim\mathcal{UF}(-\iota_\mathrm{p},\iota_\mathrm{p})$. Therefore, since $\mathbf{Z}$ and $\mathbf{h}$ are independent and $\mathbb{E}[\boldsymbol{\nu}]=\mathbb{E}[\mathbf{w}]=\mathbf{0}_T$, the mean of $\mathbf{y}$ is $\sqrt{\rho\varrho_\text{t}\varrho_\text{r}}\xi\overline{\mathbf{Z}}\mathbb{E}[\mathbf{h}]$.

The covariance of $\mathbf{h}$ is given by $\mathbf{C}_{\mathbf{h}\mathbf{h}}=\mathbb{E}[\mathbf{h}\mathbf{h}^{\text{H}}]
-\mathbb{E}[\mathbf{h}]\mathbb{E}[\mathbf{h}]^{\text{H}}$. Then, the ($n_1,n_2$)th element of $\mathbf{C}_{\mathbf{h}\mathbf{h}}$ obeys $[\mathbf{C}_{\mathbf{h}\mathbf{h}}]_{n_1,n_2}=\mathbb{E}[g_{\text{t},n_1}
g_{\text{r},n_1}^{*}g_{\text{t},n_2}^{*}g_{\text{r},n_2}]
-\overline{g}_{\text{t},n_1}\overline{g}_{\text{r},n_1}^{*}\overline{g}_{\text{t},n_2}^{*}
\overline{g}_{\text{r},n_2}$, where $\mathbb{E}[g_{\text{t},n_1}g_{\text{r},n_1}^{*}g_{\text{t},n_2}^{*}g_{\text{r},n_2}]$ equals 1 when $n_1=n_2$ and equals $\overline{g}_{\text{t},n_1}\overline{g}_{\text{r},n_1}^{*}\overline{g}_{\text{t},n_2}^{*}
\overline{g}_{\text{r},n_2}$ when $n_1\neq n_2$. Thus, we can get $[\mathbf{C}_{\mathbf{h}\mathbf{h}}]_{n_1,n_2}=1-\frac{\kappa_\text{t}\kappa_\text{r}}
{1+\kappa_\text{t}+\kappa_\text{r}+\kappa_\text{t}\kappa_\text{r}}$ when $n_1=n_2$ and $[\mathbf{C}_{\mathbf{h}\mathbf{h}}]_{n_1,n_2}=0$ when $n_1\neq n_2$. Therefore, $\mathbf{C}_{\mathbf{h}\mathbf{h}}=\mathbf{I}_N-\frac{\kappa_\text{t}\kappa_\text{r}}
{1+\kappa_\text{t}+\kappa_\text{r}+\kappa_\text{t}\kappa_\text{r}}\mathbf{I}_N=\eta\mathbf{I}_N$.

Based on (\ref{Channel_Estimation_1}), the covariance between $\mathbf{h}$ and $\mathbf{y}$ is given by
\begin{align}\label{Appendix_A_5}
    \mathbf{C}_{\mathbf{h}\mathbf{y}}\overset{(a)}{=}\sqrt{\rho\varrho_\text{t}
    \varrho_\text{r}}\mathbf{C}_{\mathbf{h}\mathbf{h}}\mathbb{E}[\mathbf{Z}^{\text{H}}]
    +\sqrt{\varrho_\text{r}}\mathbf{C}_{\mathbf{h}\boldsymbol{\nu}}
    +\mathbf{C}_{\mathbf{h}\mathbf{w}}\overset{(b)}{=}\sqrt{\rho\varrho_\text{t}
    \varrho_\text{r}}\eta\xi\overline{\mathbf{Z}}^{\text{H}},
\end{align}
where (a) is based on the independence of $\mathbf{Z}$ and $\mathbf{h}$, while (b) is based on that $\mathbf{C}_{\mathbf{h}\mathbf{h}}=\eta\mathbf{I}_N$, $\mathbb{E}[\mathbf{Z}]=\xi\overline{\mathbf{Z}}$ and $\mathbf{C}_{\mathbf{h}\boldsymbol{\nu}}=\mathbf{C}_{\mathbf{h}\mathbf{w}}=\mathbf{O}_{T}$ due to the independence of $\mathbf{h}$ and $a_{t,n}$ as well as that of $\mathbf{h}$ and $w_{t}$.

Since the thermal noise of the active reflection
amplifier $a_{t,n}$ is i.i.d for $n=1,\cdots,N$ and $t=1,\cdots,T$, we get $\mathbf{C}_{\boldsymbol{\nu}\boldsymbol{\nu}}=\varrho_\text{r}\sigma_a^2\ddot{\beta}\mathbf{I}_T$ after some further manipulations. Finally, $\mathbf{C}_{\mathbf{y}\mathbf{y}}$ in (\ref{Channel_Estimation_7}) may be arrived at based on the independence of $\mathbf{Z}$ and $\mathbf{h}$, and exploiting that $\mathbf{C}_{\mathbf{h}\mathbf{h}}=\eta\mathbf{I}_N$, $\mathbb{E}[\mathbf{Z}]=\xi\overline{\mathbf{Z}}$, $\mathbf{C}_{\boldsymbol{\nu}\boldsymbol{\nu}}=\varrho_\text{r}\sigma_a^2\ddot{\beta}\mathbf{I}_T$, the independence of $\mathbf{h}$ and $a_{t,n}$ as well as that of $\mathbf{h}$ and $w_{t}$.

\section{}\label{Appendix_B}
According to (\ref{Channel_Estimation_3}), we can express the covariance of the estimated channel as $\mathbf{C}_{\hat{\mathbf{h}}\hat{\mathbf{h}}}=\mathbf{C}_{\mathbf{h}\mathbf{y}}
\mathbf{C}_{\mathbf{y}\mathbf{y}}^{-1}\mathbf{C}_{\mathbf{h}\mathbf{y}}^{\text{H}}$, where $\mathbf{C}_{\mathbf{h}\mathbf{y}}$ and $\mathbf{C}_{\mathbf{y}\mathbf{y}}$ are given in (\ref{Channel_Estimation_6}) and (\ref{Channel_Estimation_7}) respectively. Thus, we arrive at
\begin{align}\label{Appendix_B_1}
    \notag\mathbf{C}_{\hat{\mathbf{h}}\hat{\mathbf{h}}}=&\rho\varrho_\text{t}\varrho_\text{r}
    \xi^2\eta^2\overline{\mathbf{Z}}^{\text{H}}(\rho\varrho_\text{t}\varrho_\text{r}
    (\xi^2\eta\overline{\mathbf{Z}}\overline{\mathbf{Z}}^{\text{H}}+(1-\xi^2)
    \ddot{\beta}\mathbf{I}_T)\\
    \notag&+\varrho_\text{r}\sigma_a^2\ddot{\beta}\mathbf{I}_T+\sigma_w^2\mathbf{I}_T)
    ^{-1}\overline{\mathbf{Z}}\\
    \notag=&\frac{\rho\varrho_\text{t}\varrho_\text{r}\xi^2\eta^2}{\rho\varrho_\text{t}
    \varrho_\text{r}(1-\xi^2)\ddot{\beta}+\varrho_\text{r}\sigma_a^2\ddot{\beta}
    +\sigma_w^2}\overline{\mathbf{Z}}^{\text{H}}(\mathbf{I}_T+\\
    \notag&\frac{\rho\varrho_\text{t}\varrho_\text{r}\xi^2\eta}{\rho\varrho_\text{t}
    \varrho_\text{r}(1-\xi^2)\ddot{\beta}+\varrho_\text{r}\sigma_a^2\ddot{\beta}
    +\sigma_w^2}\overline{\mathbf{Z}}\overline{\mathbf{Z}}^{\text{H}})^{-1}
    \overline{\mathbf{Z}}\\
    \notag\overset{(a)}{=}&\frac{\rho\varrho_\text{t}\varrho_\text{r}\xi^2\eta^2}
    {\rho\varrho_\text{t}\varrho_\text{r}(1-\xi^2)\ddot{\beta}+\varrho_\text{r}
    \sigma_a^2\ddot{\beta}+\sigma_w^2}\overline{\mathbf{Z}}^{\text{H}}(\mathbf{I}_T
    -\overline{\mathbf{Z}}\\
    \notag&(\overline{\mathbf{Z}}^{\text{H}}\overline{\mathbf{Z}}
    +\frac{\rho\varrho_\text{t}\varrho_\text{r}(1-\xi^2)\ddot{\beta}
    +\varrho_\text{r}\sigma_a^2\ddot{\beta}+\sigma_w^2}{\rho\varrho_\text{t}\varrho_\text{r}
    \xi^2\eta})^{-1}\overline{\mathbf{Z}}^{\text{H}})\overline{\mathbf{Z}}\\
    \overset{(b)}{=}&\text{diag}\{\varepsilon_1,\varepsilon_2,\cdots,\varepsilon_N\},
\end{align}
where (a) is based on the matrix inversion lemma that $(\mathbf{A}+\mathbf{B}\mathbf{F}\mathbf{D})^{-1}=\mathbf{A}^{-1}-\mathbf{A}^{-1}
\mathbf{B}(\mathbf{D}\mathbf{A}^{-1}\mathbf{B}+\mathbf{F}^{-1})^{-1}\mathbf{D}
\mathbf{A}^{-1}$ by letting $\mathbf{A}=\mathbf{I}_T$, $\mathbf{B}=\overline{\mathbf{Z}}$, $\mathbf{F}=\frac{\rho\varrho_\text{t}\varrho_\text{r}\xi^2\eta}{\rho\varrho_\text{t}
\varrho_\text{r}(1-\xi^2)\ddot{\beta}+\varrho_\text{r}\sigma_a^2\ddot{\beta}
+\sigma_w^2}\mathbf{I}_T$ and $\mathbf{D}=\overline{\mathbf{Z}}^{\text{H}}$. Furthermore, (b) is based on $\overline{\mathbf{Z}}^{\text{H}}\overline{\mathbf{Z}}=T\cdot\text{diag}\left\{\beta_1^2,
\cdots,\beta_N^2\right\}$, when the phase of $\overline{\mathbf{Z}}$ is designed based on the DFT or Hadamard matrix, on the definition of $\varepsilon_n$ and on some matrix manipulations.

The covariance of $\widetilde{\mathbf{h}}$ in (\ref{Channel_Estimation_10}) can be arrived at since $\mathbf{C}_{\widetilde{\mathbf{h}}\widetilde{\mathbf{h}}}=\mathbf{C}_{\mathbf{h}\mathbf{h}}
-\mathbf{C}_{\hat{\mathbf{h}}\hat{\mathbf{h}}}$, where $\mathbf{C}_{\mathbf{h}\mathbf{h}}=\eta\mathbf{I}_N$ and $\mathbf{C}_{\hat{\mathbf{h}}\hat{\mathbf{h}}}$ is given in (\ref{Channel_Estimation_8}).

\section{}\label{Appendix_C}
Based on the mean and covariance of $\hat{\mathbf{h}}$, we employ the complex normal distribution for approximating $\hat{h}_n$ as $\hat{h}_n\sim \mathcal{CN}(\overline{g}_{\text{t},n}\overline{g}_{\text{r},n}^{*},\varepsilon_n)$. Therefore, the amplitude of $\hat{h}_n$ follows the Rician distribution, denoted as $|\hat{h}_n|\sim\mathcal{RC}(\kappa_n,\Omega_n)$, where $\kappa_n=\frac{1-\eta}{\varepsilon_n}$ is the shaping parameter defined as the ratio of the power contributions by the determined component to the undetermined component. Furthermore $\Omega_n$ is the scaling parameter defined as the total power of all components, i.e. $\Omega_n=1-\eta+\varepsilon_n$. The $k$th ($k=1,2,3,4$) moments of $|\hat{h}_n|$, denoted by $\mu_n^{(k)}$, are $\mu_n^{(1)}=\sqrt{\frac{\pi\Omega_n}{4(\kappa_n+1)}}L_{\frac{1}{2}}
(-\kappa_n)$, $\mu_n^{(2)}=\Omega_n$, $\mu_n^{(3)}=\sqrt{\frac{9\pi\Omega_n^3}{16(\kappa_n+1)^3}}L_{\frac{3}{2}}
(-\kappa_n)$ and $\mu_n^{(4)}=\frac{(\kappa_n^2+4\kappa_n+2)\Omega_n^2}{(\kappa_n+1)^2}$, respectively. Since $\mathbf{C}_{\hat{\mathbf{h}}\hat{\mathbf{h}}}$ is a diagonal matrix, we arrive at:
\begin{align}\label{Appendix_C_1}
    \notag\mathbb{E}\big[(\sum\nolimits_{n=1}^{N}\beta_n|\hat{h}_n|)^2\big]
    =&\sum\nolimits_{n=1}^{N}\beta_n^2\mu_n^{(2)}+\\
    &\sum\nolimits_{(n_1,n_2)=1}^{N}\beta_{n_1}\beta_{n_2}\mu_{n_1}^{(1)}\mu_{n_2}^{(1)},
\end{align}
\begin{align}\label{Appendix_C_2}
    \notag&\mathbb{E}\big[(\sum\nolimits_{n=1}^{N}\beta_n|\hat{h}_n|)^4\big]\\
    \notag=&\binom{4}{4}\sum_{n=1}^{N}\beta_n^4\mu_n^{(4)}+\binom{4}{3,1}
    \sum\nolimits_{(n_1,n_2)=1}^{N}\beta_{n_1}^3\beta_{n_2}\mu_{n_1}^{(3)}\mu_{n_2}^{(1)}\\
    \notag&+\binom{4}{2,2}\sum_{(n_1,n_2)=1}^{N}\beta_{n_1}^2\beta_{n_2}^2
    \mu_{n_1}^{(2)}\mu_{n_2}^{(2)}\\
    \notag&+\binom{4}{1,1,1,1}\sum_{(n_1,n_2,n_3,n_4)=1}^{N}\beta_{n_1}
    \beta_{n_2}\beta_{n_3}\beta_{n_4}\mu_{n_1}^{(1)}\mu_{n_2}^{(1)}\mu_{n_3}^{(1)}\mu_{n_4}^{(1)}\\
    &+\binom{4}{2,1,1}\sum_{(n_1,n_2,n_3)=1}^{N}\beta_{n_1}^2\beta_{n_2}
    \beta_{n_3}\mu_{n_1}^{(2)}\mu_{n_2}^{(1)}\mu_{n_3}^{(1)}.
\end{align}
According to (\ref{Performance_Analysis_9}), (\ref{Appendix_C_1}), (\ref{Appendix_C_2}) and some further manipulations, we arrive at (\ref{Performance_Analysis_10}) and (\ref{Performance_Analysis_11}).

\section{}\label{Appendix_D}
According to (\ref{Performance_Analysis_10}), when $\beta=\beta_1=\beta_2=\cdots=\beta_N$ and $\varepsilon=\varepsilon_1=\varepsilon_2=\cdots=\varepsilon_N$, the mean of the received SINR can be formulated as:
\begin{align}\label{Appendix_D_1}
    \mathbb{E}[\gamma]=\gamma_0\Big[N\beta^2
    (1-\eta+\varepsilon)
    +\frac{\pi}{4}\frac{N(N-1)}{2}\beta^2
    \varepsilon L_{\frac{1}{2}}^2
    \Big(\frac{\eta-1}{\varepsilon}\Big)\Big].
\end{align}
When $N\rightarrow\infty$, we arrive at:
\begin{align}\label{Appendix_D_2}
    \notag\frac{\mathbb{E}[\gamma]}{N}=&\frac{\gamma_0\Big[N\beta^2
    (1-\eta+\varepsilon)+\frac{\pi}{4}\frac{N(N-1)}{2}\beta^2
    \varepsilon L_{\frac{1}{2}}^2
    \Big(\frac{\eta-1}{\varepsilon}\Big)\Big]}{N}\\
    \rightarrow&\frac{\pi}{4}\frac{\xi^2\varepsilon}{\eta-\xi^2\varepsilon}L_{\frac{1}{2}}^2
    \Big(\frac{\eta-1}{\varepsilon}\Big).
\end{align}
Due to the channel hardening effect when $N\rightarrow\infty$ and $\rho\rightarrow\infty$, we can get
\begin{align}\label{Appendix_D_3}
    \frac{R}{\log_2(1+\mathbb{E}[\gamma])}
    =\frac{\mathbb{E}[\log_2(1+\gamma)]}{\log_2(1+\mathbb{E}[\gamma])}\rightarrow1.
\end{align}
According to (\ref{Appendix_D_2}) and (\ref{Appendix_D_3}), we the have:
\begin{align}\label{Appendix_D_4}
    R\rightarrow\log_2\Big(1+\frac{\pi}{4}\frac{\xi^2\varepsilon}{\eta-\xi^2\varepsilon}
    L_{\frac{1}{2}}^2\Big(\frac{\eta-1}{\varepsilon}\Big)N\Big),
\end{align}
and (\ref{Ergodic_Channel_Capacity_5}) can be obtained by setting $\eta=1$.

\bibliographystyle{IEEEtran}
\bibliography{IEEEabrv,TAMS}
\end{document}